\documentclass[aps,pra,floatfix,showpacs,twocolumn,tightenlines,superscriptaddress]{revtex4}

\usepackage{amsmath}
\usepackage{graphicx}
\usepackage{epsfig}

\newcommand{\ket}[1]{\mbox{$|#1\rangle$}}
\newcommand{\bra}[1]{\mbox{$\langle #1|$}}
\newcommand{\braket}[2]{\mbox{$\langle #1|#2\rangle$}}
\newcommand{\ketbra}[2]{\mbox{$|#1\rangle\langle #2|$}}
\newcommand{\op}[1]{\mbox{\boldmath $\hat{#1}$}}

\newcommand{\binomial}[2]{\binom{#1}{#2}}

\newcommand{\pad}{PAD}

\begin{document}

\title{Photon Added Detection}

\author{A.M.~Bra{\'n}czyk}
\affiliation{Centre for Quantum Computer Technology and
Department of Physics, University of Queensland, QLD 4072, Australia}
\author{Tobias~J.~Osborne}
\affiliation{Centre for Quantum Computer Technology and
Department of Physics, University of Queensland, QLD 4072, Australia}
\affiliation{Department of Mathematics, University of Bristol, UK.}
\author{Alexei~Gilchrist}
\email{alexei@physics.uq.edu.au}
\affiliation{Centre for Quantum Computer Technology and
Department of Physics, University of Queensland, QLD 4072, Australia}
\author{T.C.~Ralph}
\affiliation{Centre for Quantum Computer Technology and
Department of Physics, University of Queensland, QLD 4072, Australia}

\date{\today}

\begin{abstract}
  The production of conditional quantum states and quantum operations
  based on the result of measurement is now seen as a key tool in
  quantum information and metrology. We propose a new type of photon
  number detector. It functions non-deterministically, but when
  successful, it has high fidelity. The detector, which makes use of
  an $n$-photon auxiliary Fock state and high efficiency Homodyne
  detection, allows a tunable tradeoff between fidelity and
  probability. By sacrificing probability of operation, an excellent
  approximation to a photon number detector is achieved.
\end{abstract}

\pacs{42.50Dv}

\maketitle

\section{Introduction}

In quantum theory, measurements encapsulate our observation of nature.
They are the link between the abstract machinery of the theory and its
observational consequences. Because of this, it is not surprising that
often new measurement techniques and strategies can drive new
applications. Moreover, the production of conditional quantum states
and quantum operations based on the results of measurement is now seen
as a key tool in realizing quantum information processing goals
\cite{klm,99gottesman390}. In optical schemes, conditional
measurements provide an effective nonlinearity that allows optical
quantum gates to be fashioned
\cite{klm,01pittman062311,02ralph012314,0110144,0110115}, and the
creation of highly entangled states suitable for quantum metrology
\cite{01lee030101,02kok052104,02fiurasek053818,02zou014102}.

Often, however, the ideal measurements envisioned in theoretical
proposals are not so easily realized experimentally. Linear optics
quantum computation schemes such as in~\cite{klm}, require high
efficiency \emph{selective} detectors (detectors able to distinguish
between zero, one and several photons). The most promising detector
candidate in this regard is the visible-light photon counter (VLPC)
\cite{99kim902,99takeuchi1063} which has achieved efficiencies of the order
of 88\%. Unfortunately these detectors require extreme operating conditions
and suffer from high dark-count rates.

\begin{figure}
  \begin{center}
   \includegraphics[width=.7\columnwidth]{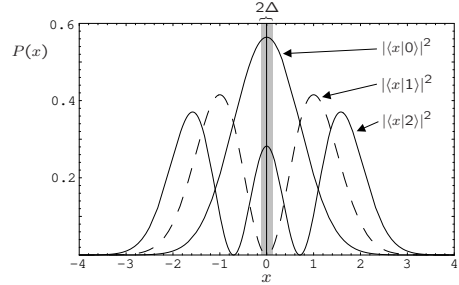}
  \end{center}
  \caption{The probability density of getting a particular $x$ value 
    if we measure the $X$ quadrature using homodyne detection. Results
    shown for various initial Fock states.}
  \label{fig:pxn}
\end{figure}

In this manuscript we introduce the idea of a non-deterministic
detector based on \emph{photon added detection} (\pad{}), where we
make use of high efficiency homodyne detection and mix the input
state with an $\ket{n}$ Fock state prior to detection. This detector
works non-deterministically, and there is an essential trade-off
between the probability that the detector works and the degree to
which the detector functions as an $n$-Fock state
projector. When the detector fails, this is clearly signalled in the
output. The essence of the detecting scheme is based on the
observation that if we use homodyne detection and post-select within a
narrow band of $2\Delta$ around $x=0$ then the detection will only be
sensitive to even photon numbers, see figure~\ref{fig:pxn}. By careful
use of quantum interference, we can make the detector act like a
projector onto a particular photon number.

The structure of the paper is as follows. First we will introduce the
scheme in general, then focus on the limiting case where $\Delta=0$ to
motivate its function. We then consider the effect of a finite
$\Delta$ and discuss the trade off between probability of operation
and fidelity.  Finally, before concluding, we examine the effect of
detector inefficiencies in our scheme.

\section{The Scheme}

In order to characterise how well the detector functions we shall
calculate the ability of the detector to pick out an appropriate state
$\ket{a_p}$ from an entangled state of the form
\begin{equation}
  \label{eq:test-state}
  \ket{\psi} =\mathcal{N}_0\sum_{n=p-w}^{p+w}\ket{a_n}_a\ket{n}_b
\end{equation}
when we measure mode $b$. The normalisation is $\mathcal{N}_0=
\frac{1}{\sqrt{2w+1}}$, and the parameter $w$ defines a window of
states, from which we want to pick out the central component.  The
reason for choosing this comparison is two-fold. Firstly we are
interested in states precisely of the above form where the states
$\ket{a_n}$ represent multi-mode states which we are conditioning by
detection and post-selection.  Secondly, this approach provides an
easily computable measure of how close to a $\ketbra{p}{p}$ projector
the detector functions in this context, since this approach reduces to
a characterisation of state preparation \cite{01kok033812}.

\begin{figure}
  \begin{center}
    \includegraphics{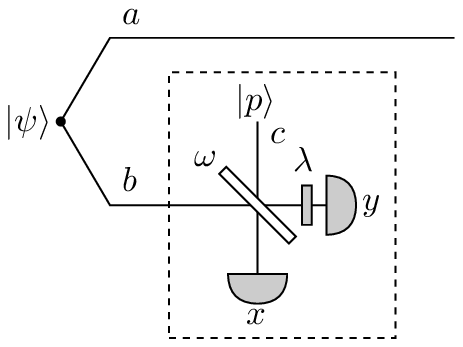}
  \end{center}
  \caption{Quantum circuit describing our detector arrangement.}
  \label{fig:det}
\end{figure}
With this characterisation in mind, consider the circuit in
figure~\ref{fig:det}. We have some multi-mode state $\ket{\psi}$ and we
wish to condition the state of mode(s) $a$ dependent on a photon
number measurement on mode $b$. For simplicity consider only a single
$n$-photon Fock state component in mode $b$, the general case is
recovered through additivity, i.e. $\ket{\psi}=\mathcal{N}_0\sum_n
\ket{\psi^{(n)}}$.  The input state is then some state
$\ket{\psi^{(n)}}=\ket{a_n}_a\ket{n}_b\ket{p}_c$, where $\ket{a_n}$ is the
associated component in mode $a$ and mode $c$ is initially in a
$p$-photon Fock state. After interacting on a beam-splitter of
reflectivity $\cos^2(\omega)$ and undergoing a phase shift $\lambda$
on mode $b$ the output state is
\begin{align}
\ket{\psi^{(n)}_{\mathrm{out}}}=&\frac{\ket{a_n}_a}{\sqrt{n!p!}}
\sum_{m=0}^{n}\sum_{q=0}^{p}\binomial{n}{m}\binomial{p}{q}e^{i\pi(p-q)+i(m+q)\lambda}\nonumber\\
&\times c^{m+p-q}s^{n-m+q}\op{b}^{\dagger m+q}\op{c}^{\dagger n+p-(m+q)}\ket{00}_{bc}
\end{align}
where $\op{b}^\dagger$ and $\op{c}^{\dagger}$ are the bosonic creation
operators for modes $b$ and $c$ respectively, $c=\cos(\omega)$,
$s=\sin(\omega)$, and finally we also have the usual binomial coefficients
$\binom{u}{v}=\frac{u!}{(u-v)!v!}$.

Modes $b$ and $c$ are now detected using separate balanced homodyne
detectors.  To an excellent approximation such detectors can be
modeled as projectors onto small ranges of quadrature amplitude
eigenstates $\ket{x_\theta}$ where $x_\theta$ is a continuous variable
with infinite dimension, and $\theta$ describes the phase relationship
with the local oscillator of the homodyne detector.   The
final conditional state (unnormalised), given we obtain $x_\theta$ in
one detector and $y_\phi$ in the other, is $\ket{\psi_\mathrm{cond}}=
\mathcal{N}_0\sum_n\ketbra{x,y}{x,y}\ket{\psi^{(n)}_\mathrm{out}}=
\mathcal{N}_0\sum_n\ket{\psi^{(n)}_\mathrm{cond}}$ where,
\begin{align}
  \ket{\psi^{(n)}_\mathrm{cond}} =& \frac{e^{-i(n+p)\phi-\frac{1}{2}(x_\theta^2+y_\phi^2)}\ket{a_n,x_\theta,y_\phi}}{\sqrt{n!p!\pi 2^{n+p}}}\nonumber\\
&\times\sum_{m=0}^{n}\sum_{q=0}^{p}\binomial{n}{m}\binomial{p}{q}e^{i\pi(p-q)+i(m+q)(\lambda-\theta+\phi)}\nonumber\\
&\times c^{m+p-q}s^{n-m+q}H_{{m+q}}(x_\theta)H_{{n+p-(m+q)}}(y_\phi)
\label{eq:psi_cond}
\end{align}
where we have used the fact that the overlap
between the quadrature amplitude eigenstates and the number states is
given by
\begin{equation}
  \braket{x_\theta}{n} = \frac{H_n(x_\theta)}{\sqrt{\sqrt{\pi}2^n n!}}
e^{-\frac{1}{2}x_\theta^2-in\theta }
\end{equation}
and $H_n(x)$ is the Hermite polynomial of order $n$. We have chosen the
convention that the $\theta=0$ quadrature operator can be written in
terms of the mode operators as $X = (a+a^\dagger)/\sqrt{2}$. Notice
that the quadrature phase angles $\theta$ and $\phi$ are effectively
not independent of $\lambda$ and that without loss of generality we
can absorb those terms into $\lambda$ (so we will take
$\lambda-\theta+\phi\rightarrow \lambda$). For simplicity we shall
also take $\phi=0$ and set the overall phase of this component to
zero, and hence we can also drop the quadrature angle subscript on $x$
and $y$.  Now consider the case where we use a 50:50 beam-splitter so
that $\omega=\pi/4$ and we set $\lambda=\pi/2$. With these conditions
equation~(\ref{eq:psi_cond}) reduces to
\begin{align}
\label{eq:psi_cond_simple}
   \ket{\psi^{(n)}_\mathrm{cond}} =&
\frac{e^{-\frac{1}{2}(x^2+y^2)}e^{i\pi p}}{\sqrt{n!p!\pi} 2^{n+p}}
g(n,p)\ket{a_n,x,y}\\
g(n,p) =&\sum_{m=0}^{n}\sum_{q=0}^{p}\binomial{n}{m}\binomial{p}{q}
e^{i\frac{\pi}{2}(m-q)}\nonumber\\
&\times H_{{m+q}}(x)H_{{n+p-(m+q)}}(y)
\end{align}

To see how this detecting scheme is only sensitive to the $p$-Fock
component we focus on the limiting case of $\Delta=0$ next.

\section{Limiting Case}

\begin{figure*}[htb]
  \begin{center}
   \includegraphics[width=.9\textwidth]{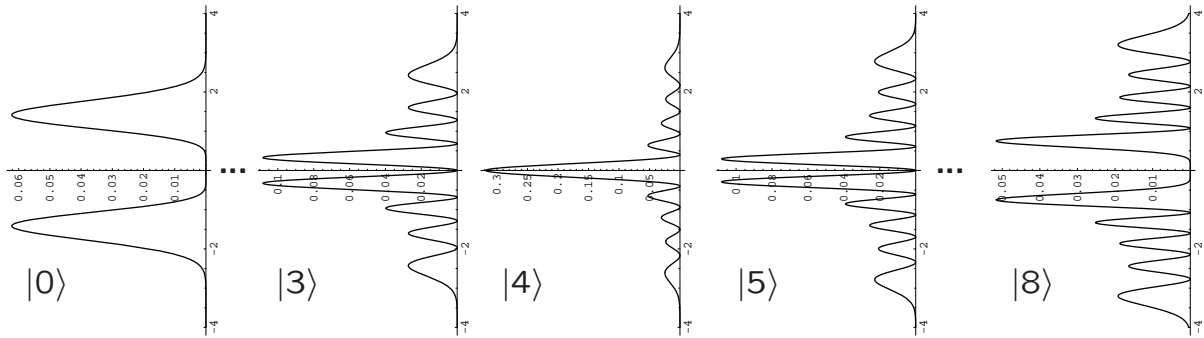}
  \end{center}
  \caption{Probability density plots for obtaining a particular $x$ and $y$ for the 
    homodyne detections given an auxiliary photon number of
    $\ket{p}=\ket{4}$, for various input number states $\ket{n}$. Only
    the $x$ axis shown as the distributions are rotationally
    symmetric.  By post-selecting on a narrow band near $x=0$ the
    detector becomes only sensitive to components with
    $\ket{n}=\ket{p}$. Also, the noise form having a finite
    post-selection band comes from the nearby number states from the
    target state. }
  \label{fig:density}
\end{figure*}

Consider only the special case where we happen to detect $x=y=0$ in
the homodyne detectors. For these values, we can use
\begin{equation}
H_n(0) = \left\{ \begin{array}{cl}0&n\;\mathrm{odd}\\ \frac{(-1)^{n/2}n!}{(n/2)!} & n\;\mathrm{even}\end{array}\right.
\end{equation}
This relation implies that only terms with even $m+q$ will be
non-zero, which in turn implies that $n+p$ must be even also.  If we
now write $g \rightarrow [g(n,p)+g'(n,p)]/2$ where $g'(n,p)$ simply
has the order of the summations reversed, we get
\begin{gather}
  g(n,p)=\frac{1}{2}\sum_{m=0}^{n}\sum_{q=0}^{p}\binomial{n}{m}\binomial{p}{q}
H_{{m+q}}(0)H_{{n+p-(m+q)}}(0)\nonumber\\
e^{i\frac{\pi}{2}(m-q)}\left( 1+e^{i\pi k} \right)
\end{gather}
where we have set $n=p+2k$ and used the fact that $m+q$ must be even.
From this expression it is clear that terms with odd $k$ will also vanish.
Terms with even $k>0$ will also vanish --- this can be readily verified
numerically. This then only leaves the terms with $k=0$ ($n=p$) as 
contributing to the state~(\ref{eq:psi_cond_simple}) and so the detector
picks out the $\ket{a_p}$ component.

This analysis assumes an infinitesimal acceptance band for the
detector. In order to assess the practicalities of the system we
need to integrate over some range of values around $x=y=0$ and
evaluate success and failure probabilities. Clearly there will be a
tradeoff between how well we project onto the $p$-photon Fock state
and the probability of obtaining a successful outcome.

\section{Finite $\Delta$}

The probability density for obtaining a value $x$ in mode $c$ and $y$
in mode $b$ will be
\begin{eqnarray}
  \label{eq:det-prob}
  P(x,y) &=& \mathrm{tr}\{ \ketbra{x}{x}\otimes\ketbra{y}{y} \rho  \}\\
  &=& \mathrm{tr}_a \{ \bra{x,y}\rho\ket{x,y} \}
\end{eqnarray}
where $\rho$ is the three mode density matrix describing the state
after the beam-splitter. This distribution is radially symmetric about
the origin, so we will switch to the polar co-ordinates $r$ and $\theta$
(where $r^2=x^2+y^2$) and accept a particular result if it lies within
a certain radius $\Delta$. Intuitively we can see what the effect will
be from figure~\ref{fig:density}. As we make $\Delta$ larger, the
probability that a result falls within the accepted band, picks up
contributions from nearby states to the target state, and these will
contribute to the error.  The total probability that we get $0 \le r
\le \Delta$ is
\begin{equation}
  \label{eq:Pdelta}
  P_\Delta = 2\pi \int_{0}^{\Delta} P(r,\theta) r dr
\end{equation}
The (unnormalised) state immediately after destructively obtaining a
particular $x$ and $y$ in the first two modes is
$  \rho^{(x,y)}_a =  \bra{x,y}\rho\ket{x,y}$.
Consequently the ensemble of states that we would obtain if we where
to only accept values within a radius $\Delta$, would be
\begin{equation}
  \label{eq:rho-out}
  \rho_a = \frac{1}{P_\Delta}  \int_{0}^{2\pi} d\theta 
\int_{0}^{\Delta} dr \; \rho^{(r,\theta)}_a
\end{equation}

\begin{figure}[htb]
  \begin{center}
   \includegraphics[width=.9\columnwidth]{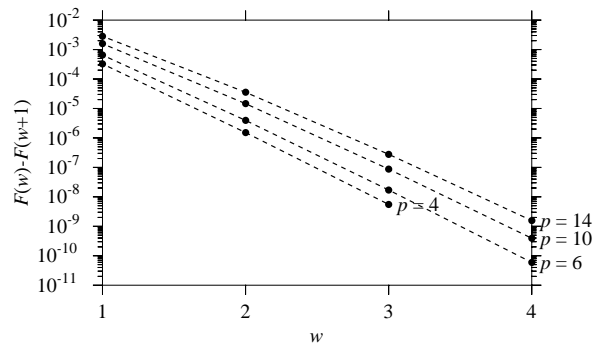}
  \end{center}
  \caption{The main source of error for the detector is due to contributions 
    from number states near the target state. Here we plot the difference in fidelity
    $F$, between two successive window sizes, $w$ and $w+1$. As can be seen, increasing the size of
    the window of states we are testing against makes little
    difference past a few states, consequently we will adopt $w=2$ in calculations. Note that $\Delta=0.1$
    in the plots.}
  \label{fig:win-converge}
\end{figure}

To compare how well such a projector functions we can use the fidelity
against the target state $\ket{a_p}$:
\begin{equation}
  \label{eq:det-fid}
  F(\Delta) = |\bra{a_p}\rho_a\ket{a_p} | 
\end{equation}
Note that in calculating this quantity we will assume that the
$\ket{a_j}$ are orthonormal.

One of the important features of the \pad{} scheme is that it is
sensitive only to a band of number states near the target state. This
effect can be seen in the behaviour of the probability densities for
states far away from the target state in figure~\ref{fig:density}, and
is clearly demonstrated in figure~\ref{fig:win-converge}, where we
show the rapid convergence in fidelity as we increase the number of
nearby states to the one we are projecting out.

\begin{figure}[htb]
  \begin{center}
   \includegraphics[width=.9\columnwidth]{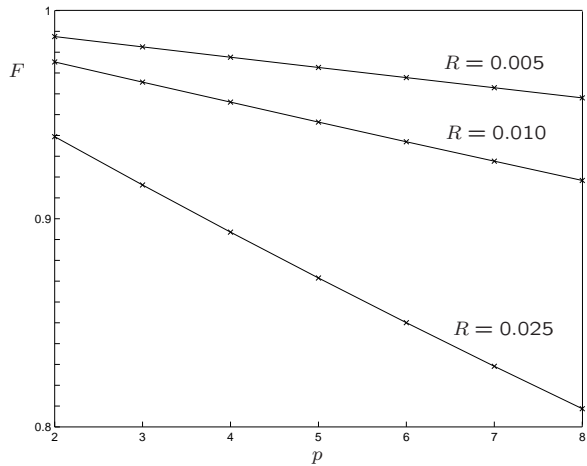}
  \end{center}
  \caption{The fidelity of operation for various target states $\ket{a_p}$ 
    from the distributions (\ref{eq:test-state}) is given (with $w=2$). The curves
    are for fixed probability rates $R$.}
  \label{fig:rates}
\end{figure}

As we increase $\Delta$, the probability that we get a result we will
accept also increases, but due to the overlap with the states near the
target state the fidelity of the detector will drop. The actual
probability is not a meaningful quantity in this context as it depends
as much on the test state (\ref{eq:test-state}) as on the parameters
of the detector.  The quantity we will use instead is a probability
rate $R=P_\Delta/P_\mathrm{ideal}$, which is the probability we get
divided by the expected probability if we had an ideal photo-counter.
The tradeoff between fidelity and probability is
quantified in figure~\ref{fig:rates}.

\section{Inefficient Detection}

The calculations so far have assumed unit efficiency detection.  In
this section we explore the effect of non-unit detection efficiencies
for the \pad{}, although it should be noted from the outset that
detection efficiency for homodyne detection is very high (in the
region of 98\% \cite{92polzik3020}).  We will compare the performance
of the \pad{} to an ideal, but inefficient photon counter, which we
model by the POVM elements $\Pi_p: p=0,1,\ldots$, where $p$ is the
number of detected photons, with
\begin{equation}
  \Pi_p = \sum_{m=p}^\infty \binomial{m}{p} \eta^p(1-\eta)^{m-p}\ketbra{m}{m}
\end{equation}
Visible-light photon counters can be modelled as ideal, but
inefficient photon counters, at least for small photon numbers
\cite{0204073}.

The fidelity of the ideal detector in picking out the state $\ket{a_p}$
when used with the input state~(\ref{eq:test-state}) is then
\begin{align}
  F_\mathrm{ideal} &=\frac{ |\bra{a_p} \mathrm{Tr}_b\{\Pi_p \rho_\mathrm{in}\} \ket{a_p}|}
{ \mathrm{Tr}\{\Pi_p \rho_\mathrm{in}\}} \nonumber \\
&= \left(\sum_{n=p}^{n_{\mathrm{max}}}\binom{n}{p}(1-\eta)^{n-p}\right)^{-1}
\end{align}
where the summation extends to the maximum photon number, so for the test
state in (\ref{eq:test-state}) $n_{\mathrm{max}}=p+w$.

For the \pad{} detector we can model inefficiencies simply by
considering a beam splitter of transitivity $\eta$ in front of both
homodyne detectors \cite{93leonhardt4598}. The first observation we
make is that for high efficiency, the ideal detector obtains a higher
fidelity. The trend with higher photon number is similar for both
detectors. Where the advantage lies for the \pad{} is that the
efficiency for current homodyne detectors is very high compared with
available photon counters.

\begin{figure}[htb]
  \begin{center}
   \includegraphics[width=.9\columnwidth]{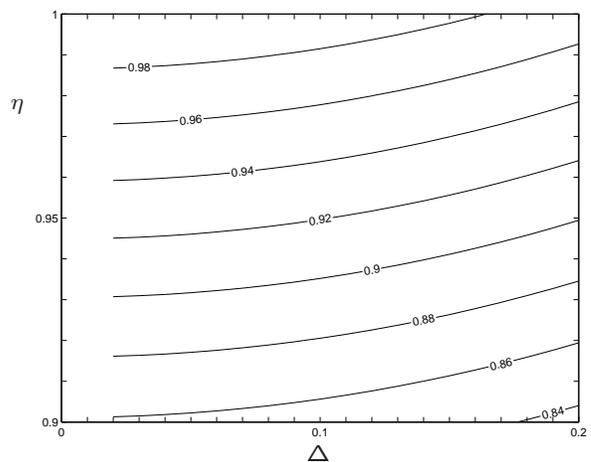}
  \end{center}
  \caption{Equivalent ideal single photon detector efficiency as a function 
    of the acceptance width $\Delta$, and the Homodyne efficiency
    $\eta$. }
  \label{fig:equic}
\end{figure}

For a particular $\Delta$ and $\eta$ we can consider an equivalent
ideal detector that gives the same fidelity. Constructing an
equivalence in this fashion is particularly useful and was considered
by \cite{02nemoto032306} where they compared an ideal photon counter
with homodyne detection in the context of quantum communication.  As
such, they used the mutual information as a means of comparison.  For
our scheme, we envision state preparation as the main application so
we will use the fidelity as a means of comparison.  This comparison is
plotted in figure~\ref{fig:equic}, for the ability to project out the
state $\ket{a_1}$ from the input state $\sum_{n=0}^4\ket{a_n}\ket{n}$.
A detector able to achieve this projection forms a \emph{selective}
detector which is needed in many linear optics schemes.

\section{Discussion and Conclusions}

Because of it's non-deterministic nature, we envision applications of
this detector mainly in state preparation, where non-classical states
are prepared through conditioning on photon number detection.  We
could prepare a good approximation to an $\ket{n}$ photon state
required by our detector, by using spontaneous parametric down
conversion and a detector cascade in one arm. Even if the detectors in
the cascade are inefficient, if, say three detectors register a click,
then we have at least a three photon term in the other arm. The errors
caused by having more than the required number of photons are offset
by the low probability of such events.  One intriguing possibility is
to employ this detector in a proposal by Dakna, \emph{et al.}
\cite{97dakna3184}.  In the Dakna scheme, a good approximation to an
optical Schr\"odinder cat state is generated by mixing a single mode
squeezed state on a beam-splitter with the vacuum and conditioning on
detecting a certain number of photons in one of the exit ports.

Another possible extension is to use other parameter choices, and
post-selection choices to directly project out certain distributions of
photon number terms.

We have presented a non-deterministic scheme which functions as a
high-fidelity Fock state projector. This detecting scheme allows a
tunable tradeoff between the fidelity and probability of detection.
The weaknesses of the scheme are that it requires an $\ket{n}$ photon
state and that it is non-deterministic. The $\ket{n}$ photon state could be
prepared in the first instance simply by conditioning the output of a
spontaneous parametric down converter with a traditional detector
cascade. The non-deterministic nature of the scheme leads us to
conclude that the main application for the detector will be in state
generation.

AG acknowledges support form the New Zealand Foundation for Research,
Science and Technology under grant UQSL0001. This project was
supported by the Australian Research Council.

\end{document}